\documentclass[preprint2]{emulateapj}
\usepackage{color}
\usepackage{amsmath}
\usepackage{threeparttable}
\usepackage{natbib}
\usepackage{times}
\usepackage{graphicx}
\usepackage{cancel}
\usepackage[mediumspace,Gray,squaren]{SIunits} 

\usepackage{comment}

\DeclareGraphicsExtensions{.jpg,.pdf,.png,.eps,.ps}

\def\muK{\rm $\mu${\mbox{K}}}

\newcommand{\degs}{deg$^2$}

\def\msun{{\rm M}_{\odot}}

\def\oz#1{{}}

\def\ell{l}

\def\Penn{1}
\def\AAUChicago{2}
\def\KICPChicago{3}
\def\StanfordKPAC{4}
\def\StanfordPhys{5}
\def\PhysicsUChicago{6}
\def\FNAL{7}
\def\UChicago{8}
\def\SLAC{9}
\def\CfA{10}
\def\MIT{11}
\def\Harvard{12}
\def\Argonne{13}
\def\Munich{14}
\def\ExcellenceCluster{15}
\def\Miss{16}
\def\EFIChicago{17}
\def\NIST{18}
\def\PUC{19}
\def\Caltech{20}
\def\McGill{21}
\def\Berkeley{22}
\def\CIFAR{23}
\def\illast{24}
\def\illphy{25}
\def\MPE{26}
\def\UFlorida{27}
\def\Colorado{28}
\def\LeidenObservatory{29}
\def\Davis{30}
\def\LBNL{31}
\def\Arizona{32}
\def\Michigan{33}
\def\Minnesota{34}
\def\Melbourne{35}
\def\STScI{36}
\def\CaseWestern{37}
\def\ArtInstChicago{38}
\def\KASI{39}
\def\LLNL{40}
\def\Dunlap{41}
\def\Toronto{42}
\def\BCCP{43}
\def\CTIO{44}

\begin{document}

\title{A measurement of gravitational lensing of the Cosmic Microwave
  Background by galaxy clusters using data from the South Pole
  Telescope}

\slugcomment{}

\author{
  \vspace{-0.02in}
  E.~J.~Baxter\altaffilmark{\Penn,\AAUChicago,\KICPChicago},
  R.~Keisler\altaffilmark{\KICPChicago,\StanfordKPAC,\StanfordPhys,\PhysicsUChicago},
  S.~Dodelson\altaffilmark{\AAUChicago,\KICPChicago,\FNAL},
  K.~A.~Aird\altaffilmark{\UChicago},
  S.~W.~Allen\altaffilmark{\StanfordKPAC,\StanfordPhys,\SLAC},
  M.~L.~N.~Ashby\altaffilmark{\CfA},
  M.~Bautz\altaffilmark{\MIT},
  M.~Bayliss\altaffilmark{\CfA,\Harvard},
  B.~A.~Benson\altaffilmark{\AAUChicago,\KICPChicago,\FNAL},
  L.~E.~Bleem\altaffilmark{\KICPChicago,\PhysicsUChicago,\Argonne},
  S.~Bocquet\altaffilmark{\Munich,\ExcellenceCluster},
  M.~Brodwin\altaffilmark{\Miss},
  J.~E.~Carlstrom\altaffilmark{\AAUChicago,\KICPChicago,\PhysicsUChicago,\Argonne,\EFIChicago},
  C.~L.~Chang\altaffilmark{\KICPChicago,\Argonne,\EFIChicago},
  I.~Chiu\altaffilmark{\Munich,\ExcellenceCluster},
  H-M.~Cho\altaffilmark{\NIST},
  A.~Clocchiatti\altaffilmark{\PUC},
  T.~M.~Crawford\altaffilmark{\AAUChicago,\KICPChicago}
  A.~T.~Crites\altaffilmark{\AAUChicago,\KICPChicago,\Caltech},
  S.~Desai\altaffilmark{\Munich,\ExcellenceCluster},
  J.~P.~Dietrich\altaffilmark{\Munich,\ExcellenceCluster},
  T.~de~Haan\altaffilmark{\McGill,\Berkeley},
  M.~A.~Dobbs\altaffilmark{\McGill,\CIFAR},
  R.~J.~Foley\altaffilmark{\illast,\illphy},
  W.~R.~Forman\altaffilmark{\CfA},
  E.~M.~George\altaffilmark{\Berkeley,\MPE},
  M.~D.~Gladders\altaffilmark{\AAUChicago,\KICPChicago},
  A.~H.~Gonzalez\altaffilmark{\UFlorida},
  N.~W.~Halverson\altaffilmark{\Colorado},
  N.~L.~Harrington\altaffilmark{\Berkeley},
  C.~Hennig\altaffilmark{\Munich,\ExcellenceCluster},
  H.~Hoekstra\altaffilmark{\LeidenObservatory},
  G.~P.~Holder\altaffilmark{\McGill},
  W.~L.~Holzapfel\altaffilmark{\Berkeley},
  Z.~Hou\altaffilmark{\KICPChicago,\PhysicsUChicago},
  J.~D.~Hrubes\altaffilmark{\UChicago},
  C.~Jones\altaffilmark{\CfA},
  L.~Knox\altaffilmark{\Davis},
  A.~T.~Lee\altaffilmark{\Berkeley,\LBNL},
  E.~M.~Leitch\altaffilmark{\AAUChicago,\KICPChicago},
  J.~Liu\altaffilmark{\Munich,\ExcellenceCluster},
  M.~Lueker\altaffilmark{\Caltech,\Berkeley},
  D.~Luong-Van\altaffilmark{\UChicago},
  A.~Mantz\altaffilmark{\KICPChicago},
  D.~P.~Marrone\altaffilmark{\Arizona},
  M.~McDonald\altaffilmark{\MIT},
  J.~J.~McMahon\altaffilmark{\Michigan},
  S.~S.~Meyer\altaffilmark{\AAUChicago,\KICPChicago,\PhysicsUChicago,\EFIChicago},
  M.~Millea\altaffilmark{\Davis},
  L.~M.~Mocanu\altaffilmark{\AAUChicago,\KICPChicago},
  S.~S.~Murray\altaffilmark{\CfA},
  S.~Padin\altaffilmark{\AAUChicago,\KICPChicago,\Caltech},
  C.~Pryke\altaffilmark{\Minnesota},
  C.~L.~Reichardt\altaffilmark{\Berkeley,\Melbourne},
  A.~Rest\altaffilmark{\STScI},
  J.~E.~Ruhl\altaffilmark{\CaseWestern},
  B.~R.~Saliwanchik\altaffilmark{\CaseWestern},
  A.~Saro\altaffilmark{\Munich},
  J.~T.~Sayre\altaffilmark{\CaseWestern},
  K.~K.~Schaffer\altaffilmark{\KICPChicago,\EFIChicago,\ArtInstChicago},
  E.~Shirokoff\altaffilmark{\AAUChicago,\KICPChicago},
  J.~Song\altaffilmark{\Michigan,\KASI},
  H.~G.~Spieler\altaffilmark{\LBNL},
  B.~Stalder\altaffilmark{\CfA},
  S.~A.~Stanford\altaffilmark{\Davis,\LLNL},
  Z.~Staniszewski\altaffilmark{\CaseWestern},
  A.~A.~Stark\altaffilmark{\CfA},
  K.~T.~Story\altaffilmark{\KICPChicago,\PhysicsUChicago},
  A.~van~Engelen\altaffilmark{\McGill},
  K.~Vanderlinde\altaffilmark{\Dunlap,\Toronto},
  J.~D.~Vieira\altaffilmark{\illast,\illphy},
  A. Vikhlinin\altaffilmark{\CfA},
  R.~Williamson\altaffilmark{\AAUChicago,\KICPChicago,\Caltech},
  O.~Zahn\altaffilmark{\BCCP},
  and 
  A.~Zenteno\altaffilmark{\Munich,\CTIO}
}

\altaffiltext{\Penn}{Center for Particle Cosmology, Department of
  Physics and Astronomy, University of Pennsylvania, Philadelphia, PA,
  USA 19104} 
\altaffiltext{\AAUChicago}{Department of Astronomy and
  Astrophysics, University of Chicago, Chicago, IL, USA 60637}
\altaffiltext{\KICPChicago}{Kavli Institute for Cosmological Physics,
  University of Chicago, Chicago, IL, USA 60637}
\altaffiltext{\StanfordKPAC}{Kavli Institute for Particle Astrophysics
  and Cosmology, Stanford University, 452 Lomita Mall, Stanford, CA
  94305} \altaffiltext{\StanfordPhys}{Department of Physics, Stanford
  University, 382 Via Pueblo Mall, Stanford, CA 94305}
\altaffiltext{\PhysicsUChicago}{Department of Physics, University of
  Chicago, Chicago, IL, USA 60637} \altaffiltext{\FNAL}{Fermi National
  Accelerator Laboratory, Batavia, IL 60510-0500, USA}
\altaffiltext{\UChicago}{University of Chicago, Chicago, IL, USA
  60637} \altaffiltext{\SLAC}{SLAC National Accelerator Laboratory,
  2575 Sand Hill Road, Menlo Park, CA 94025}
\altaffiltext{\CfA}{Harvard-Smithsonian Center for Astrophysics, Cambridge, MA, USA 02138}
\altaffiltext{\MIT}{Kavli Institute for Astrophysics and Space
Research, Massachusetts Institute of Technology, 77 Massachusetts Avenue,
Cambridge, MA 02139}
\altaffiltext{\Harvard}{Department of Physics, Harvard University, 17 Oxford Street, Cambridge, MA 02138}
\altaffiltext{\Argonne}{Argonne National Laboratory, Argonne, IL, USA 60439}
\altaffiltext{\Munich}{Department of Physics, Ludwig-Maximilians-Universit\"{a}t, 81679 M\"{u}nchen, Germany}
\altaffiltext{\ExcellenceCluster}{Excellence Cluster Universe, 85748 Garching, Germany}
\altaffiltext{\Miss}{Department of Physics and Astronomy, University of Missouri, 5110 Rockhill Road, Kansas City, MO 64110}
\altaffiltext{\EFIChicago}{Enrico Fermi Institute, University of Chicago, Chicago, IL, USA 60637}
\altaffiltext{\NIST}{NIST Quantum Devices Group, Boulder, CO, USA 80305}
\altaffiltext{\PUC}{Departamento de Astronomia y Astrosifica, Pontificia Universidad Catolica,Chile}
\altaffiltext{\Caltech}{California Institute of Technology, Pasadena, CA, USA 91125}
\altaffiltext{\McGill}{Department of Physics, McGill University, Montreal, Quebec H3A 2T8, Canada}
\altaffiltext{\Berkeley}{Department of Physics, University of California, Berkeley, CA, USA 94720}
\altaffiltext{\CIFAR}{Canadian Institute for Advanced Research, CIFAR Program in Cosmology and Gravity, Toronto, ON, M5G 1Z8, Canada}
\altaffiltext{\illast}{Astronomy Department, University of Illinois at Urbana-Champaign, 1002 W.\ Green Street, Urbana, IL 61801, USA}
\altaffiltext{\illphy}{Department of Physics, University of Illinois Urbana-Champaign, 1110 W.\ Green Street, Urbana, IL 61801, USA}
\altaffiltext{\MPE}{Max-Planck-Institut f\"{u}r extraterrestrische Physik, 85748 Garching, Germany}
\altaffiltext{\UFlorida}{Department of Astronomy, University of Florida, Gainesville, FL 32611}
\altaffiltext{\Colorado}{Department of Astrophysical and Planetary Sciences and Department of Physics, University of Colorado, Boulder, CO, USA 80309}
\altaffiltext{\LeidenObservatory}{Leiden Observatory, Leiden University, Niels Bohrweg 2, 2333 CA, Leiden, the Netherlands}
\altaffiltext{\Davis}{Department of Physics, University of California, Davis, CA, USA 95616}
\altaffiltext{\LBNL}{Physics Division, Lawrence Berkeley National Laboratory, Berkeley, CA, USA 94720}
\altaffiltext{\Arizona}{Steward Observatory, University of Arizona, 933 North Cherry Avenue, Tucson, AZ 85721}
\altaffiltext{\Michigan}{Department of Physics, University of Michigan, Ann  Arbor, MI, USA 48109}
\altaffiltext{\Minnesota}{Department of Physics, University of Minnesota, Minneapolis, MN, USA 55455}
\altaffiltext{\Melbourne}{School of Physics, University of Melbourne, Parkville, VIC 3010, Australia}
\altaffiltext{\STScI}{Space Telescope Science Institute, 3700 San Martin
Dr., Baltimore, MD 21218}
\altaffiltext{\CaseWestern}{Physics Department, Center for Education and Research in Cosmology and Astrophysics, Case Western Reserve University,Cleveland, OH, USA 44106}
\altaffiltext{\ArtInstChicago}{Liberal Arts Department, School of the Art Institute of Chicago, Chicago, IL, USA 60603}
\altaffiltext{\KASI}{Korea Astronomy and Space Science Institute, Daejeon 305-348, Republic of Korea}
\altaffiltext{\LLNL}{Institute of Geophysics and Planetary Physics, Lawrence
Livermore National Laboratory, Livermore, CA 94551}
\altaffiltext{\Dunlap}{Dunlap Institute for Astronomy \& Astrophysics, University of Toronto, 50 St George St, Toronto, ON, M5S 3H4, Canada}
\altaffiltext{\Toronto}{Department of Astronomy \& Astrophysics, University of Toronto, 50 St George St, Toronto, ON, M5S 3H4, Canada}
\altaffiltext{\BCCP}{Berkeley Center for Cosmological Physics, Department of Physics, University of California, and Lawrence Berkeley National Laboratory, Berkeley, CA, USA 94720}
\altaffiltext{\CTIO}{Cerro Tololo Inter-American Observatory, Casilla 603, La Serena, Chile}

\begin{abstract}
Clusters of galaxies are expected to gravitationally lens the cosmic
microwave background (CMB) and thereby generate a distinct signal in
the CMB on arcminute scales.  Measurements of this effect can be used
to constrain the masses of galaxy clusters with CMB data alone.  Here
we present a measurement of lensing of the CMB by galaxy clusters
using data from the South Pole Telescope (SPT).  We develop a maximum
likelihood approach to extract the CMB cluster lensing signal and
validate the method on mock data.  We quantify the effects on our
analysis of several potential sources of systematic error and find
that they generally act to reduce the best-fit cluster mass.  It is
estimated that this bias to lower cluster mass is roughly $0.85\sigma$
in units of the statistical error bar, although this estimate should
be viewed as an upper limit.  We apply our maximum likelihood
technique to 513 clusters selected via their SZ signatures in SPT
data, and rule out the null hypothesis of no lensing at
$3.1\sigma$. The lensing-derived mass estimate for the full cluster
sample is consistent with that inferred from the SZ flux:
$M_{200,\mathrm{lens}} = 0.83_{-0.37}^{+0.38}\, M_{200,\mathrm{SZ}}$
({\rm 68\%} \,C.L., statistical error only).
\end{abstract}


\section{Introduction}

\setcounter{footnote}{0}

Gravitational lensing of the cosmic microwave background (CMB) by
large-scale structure (LSS) has recently emerged as a powerful
cosmological probe.  The first detection of this effect relied on
measuring the cross-correlation between CMB lensing maps and radio
galaxy counts \citep{smith07}.  Subsequent studies have correlated CMB
lensing maps with several different galaxy populations
\citep[e.g.,][]{hirata08,bleem12b,planck13-17}, quasars
\citep[e.g.,][]{hirata08,sherwin12,planck13-17}, and maps of the
cosmic infrared background \citep{holder13,planck13-18}, to give just
a few examples.  These measurements of the correlation between CMB
lensing and intervening structure have used massive objects as
effectively point-like tracers of LSS and have thus been sensitive to
the clustering of the dark matter halos these objects inhabit.  In the
context of the halo model, this clustering signal is the ``two-halo
term'' \citep[for a review of the halo model see][]{cooray02}.

The lensing of the CMB due to the galaxies or clusters themselves is
sensitive to the structure of the individual halos, i.e.,~the
``one-halo'' term.  \citet{madhavacheril14} have recently reported
a measurement of the lensing of the CMB by dark matter halos with
masses $M \sim 10^{13}\,\msun$ using CMB data from the Atacama
Cosmology Telescope Polarimeter stacked on the locations of roughly
12,000 CMASS galaxies from the SDSS-III/BOSS survey.  Galaxy clusters,
with halo masses $M \gtrsim 10^{14}\,\msun$, offer another promising
target for measuring lensing of the CMB by individual halos.

\citet{Seljak:2000} showed that lensing by galaxy clusters induces a
dipole-like distortion in the CMB that is proportional to and aligned
with the CMB gradient behind the cluster.  Consider a galaxy cluster
lying along the line of sight to a pure gradient in the CMB.  Photon
trajectories on either side of the cluster are bent towards the
cluster, causing these photons to appear to have originated farther
away from the cluster.  The net result is that the CMB temperature
appears decreased on the hot side of the cluster and increased on the
opposite side. In the absence of a CMB temperature gradient behind the
cluster, gravitational lensing does not lead to a measurable
distortion (this can be seen as a consequence of the fact that
gravitational lensing conserves surface brightness).  The magnitude of
the CMB cluster lensing distortion is therefore sensitive to the mass
distribution of the cluster, its redshift, and also the pattern of the
CMB on the last scattering surface in the direction of the cluster.
For a typical CMB gradient of $13 \mu \rm{K}/\rm{arcmin}$ and a
cluster with mass $M \sim 10^{15}\,\msun$ located at $z \sim 1$ (a
high mass, high redshift cluster), the lensing distortion in the CMB
peaks at $\sim 10$\muK\, roughly $1$ arcminute from the cluster
center.

Current CMB experiments do not have the sensitivity to obtain high
significance detections of the lensing effect around single clusters.
To detect this effect, then, we must combine the constraints from many
clusters to increase the signal-to-noise.  Since the lensing
distortion induced by a cluster is sensitive to the mass of the
cluster, the combined lensing constraint can be translated into a
constraint on the weighted average of the cluster masses in the
sample.  For the time being, CMB lensing constraints on cluster mass
are unlikely to be competitive with other means of measuring cluster
masses, such as lensing of the light from background galaxies
\citep[e.g.,][]{johnston07, Okabe:2010, hoekstra12, high12,
  vonderlinden14a}.  Still, such measurements provide a useful
cross-check on other techniques for measuring cluster mass because
they are sensitive to different sources of systematic error.  Future
CMB experiments with higher sensitivity will dramatically improve the
signal-to-noise of CMB cluster lensing measurements.  If sources of
systematic error can be controlled, high signal-to-noise measurements
of CMB cluster lensing can provide cosmologically useful cluster mass
constraints, especially at $z \gtrsim 1$ \citep{LewisKing:2006}.
Furthermore, if both CMB lensing and galaxy lensing constraints can be
obtained on a set of clusters, these measurements can be combined to
yield interesting constraints on e.g.,~dark energy
\citep{HuHolz:2007}.

Several authors have considered the detectability of the effect and
how well CMB cluster lensing can constrain cluster masses
\citep[e.g.,][]{Seljak:2000,Holder:2004,Vale:2004,Dodelson:2004,
  LewisKing:2006, lewis06}.  Various approaches to extract the signal
have also been investigated: \citet{Seljak:2000} and \citet{Vale:2004}
considered fitting out the gradient in the CMB to extract the cluster
signal; \citet{Holder:2004} considered an approach based on Wiener
filtering; \citet{lewis06} and \citet{Yoo:2008} developed a maximum
likelihood approach; and \citet{Hu:2007} and \citet{Melin:2014}
considered approaches based on the optimal quadratic estimator of
\citet{hu01b} and \citet{hu02a}.  Many of these techniques rely on a
separation of scales inherent to the problem: the distortions caused
by cluster lensing are a few arcminutes in angular size, while the
primordial CMB has little structure on these scales as a result of
diffusion damping.  This simple picture is complicated by the fact
that instrumental noise and foreground emission may lead to arcminute
size structure in the observed temperature field.  Furthermore, any
method to extract the CMB cluster lensing signal must be robust to
contamination from the thermal and kinematic Sunyaev-Zel'dovich (SZ)
effects \citep{sunyaev72, sunyaev80}, as well as other foregrounds.

In this paper we present a $3.1\sigma$ measurement of the arcminute
scale gravitational lensing of the CMB by galaxy clusters using data
from the full 2500 \degs~SPT-SZ survey \citep[e.g.,][]{story13}.  We
develop a maximum likelihood approach to extract the CMB cluster
lensing signal based on a model for the lensing-induced distortion.
Our approach differs somewhat from those mentioned above in that it is
inherently {\it parametric}: we directly constrain the parameters of
an assumed mass profile rather than generating a map of the lensing
mass.  The method is validated via application to mock data and is
then applied to observations of the CMB around 513 clusters identified
in the SPT-SZ survey via their SZ effect signature \citep{bleem15b}.
The mass constraints from each cluster are combined to constrain the
weighted average of the cluster masses in our sample.  As a null test,
we also analyze many sets of off-cluster observations and find no
significant detection.

The paper is organized as follows: in \S\ref{sec:data} we describe the
data set used in this work and in \S\ref{sec:analysis} we develop a
maximum likelihood approach to extract the CMB cluster lensing signal
from this data set.  The results of our analysis applied to mock data
and our estimation of systematic effects are presented in
\S\ref{sec:mock_results}.  The analysis is applied to SPT data in
\S\ref{sec:spt}, and conclusions are given in \S\ref{sec:conclusion}.

\section{Data}
\label{sec:data}

\subsection{CMB Data} 
\label{subsec:cmbdata}

The data used in this work were collected with the South Pole
Telescope \citep[SPT,][]{carlstrom11} as part of the SPT-SZ survey.
The SPT-SZ survey covered roughly 2500 \degs~of the southern sky to an
approximate depth of 40, 18, and 80 \muK-arcmin in frequency bands
centered at 95, 150, and 220 GHz, respectively.  The SPT-SZ maps used
in this analysis are identical to those described in \citet{george14}.
The maps are projected using the oblique Lambert azimuthal equal-area
projection and are divided into square pixels measuring 0.5 arcminutes
on a side.

The 2500 \degs~ SPT-SZ survey area was subdivided into 19 contiguous
fields, each of which was observed to full survey depth before moving
on to the next.  The fields were observed using a sequence of
left-going and right-going scans.  Each pair of scans is at a constant
elevation, and the elevation is increased in a discrete step between
pairs.  Denoting left-going and right-going scans as $L$ and $R$, the
sky map is the sum $\frac{1}{2}(L + R)$ of maps generated from these
two scan directions.  The difference map formed via the combination
$\frac{1}{2}(L-R)$ should have no sky signal and can be used as a
statistically representative estimate of the instrumental and
atmospheric noise (henceforth, we will sometimes refer to these two
noise sources simply as ``instrumental noise,'' since the distinction
is irrelevant for our purposes).  Because the observing strategy
varies somewhat between different fields, so does the level of
instrumental noise.  Below, we will estimate the instrumental noise
levels in a field-dependent fashion.  More detailed descriptions of
the SPT observation strategy may be found in \citet{george14} and
references therein.

Each sky map used in this work is the sum of signal from the sky and
instrumental noise.  The signal contribution to the maps can be
expressed as the convolution of the true sky with an
instrumental-plus-analysis response function. The response function
characterizes how astrophysical objects would appear in the SPT-SZ
maps and consists of two components: a ``beam function'' that accounts
for the SPT beam shape, and a ``transfer function'' that accounts for
the time-stream filtering of the SPT data.  As with the instrumental
noise, variations in the observation strategy between different fields
cause the transfer function of the maps to also vary between fields.
The characterizations of the SPT transfer and beam functions are
described in \citet{george14} and references therein.  We treat the
transfer function in a field-dependent fashion below.  In
\S\ref{sec:analysis} we use the measured beam and transfer functions
to fit for the CMB cluster lensing signal in the SPT-SZ data.

\subsection{tSZ-free Maps}
\label{subsec:tszfree}

The Sunyaev-Zel'dovich effect is the distortion of the CMB induced by
inverse-Compton scattering of CMB photons and energetic electrons
\citep[for a review see][]{Birkinshaw:1999}.  This effect is
especially pronounced in the directions of massive galaxy clusters as
these objects are reservoirs of hot, ionized gas.  The SZ effect from
clusters can be divided into two parts: the thermal SZ effect (tSZ)
and the kinematic SZ effect (kSZ). The tSZ effect is due to
inverse-Compton scattering of CMB photons with hot intra-cluster
electrons.  The effect has a distinct spectral signature that makes a
cluster appear as a cold spot in the CMB at low frequencies and a hot
spot at high frequencies, with a null at 217 GHz.  If the cluster also
has a peculiar velocity relative to the CMB rest frame, the CMB will
appear anisotropic to the cluster, and an additional Doppler shift
will be imprinted on the scattered CMB photons.  This distortion,
known as the kSZ effect, is frequency independent when expressed as a
brightness temperature fluctuation.

The magnitude of the tSZ effect around galaxy clusters can be
significantly greater than the magnitude of the CMB cluster lensing
signal.  A cluster with mass $M \sim 5\times 10^{14}\,\msun$
introduces a tSZ signal of roughly $-400$\muK\, (as compared to
roughly 5\muK\, from lensing) at the cluster center when observed at
150 GHz.  Introducing this level of SZ contamination into our mock
analysis (see \S\ref{subsec:mock_data}) biases the lensing mass
constraints to such an extreme degree that we lose the ability to
measure CMB cluster lensing.  Eliminating the tSZ is therefore
essential to our analysis.

We exploit the frequency dependence of the tSZ to remove it from our
data.  Since SPT observes at 95, 150, and 220 GHz, we form a linear
combination of the data at these three frequencies that nulls
the tSZ effect, but preserves the CMB signal.  This tSZ-free linear
combination is created as follows. First, all three maps are smoothed
to the resolution of the 95 GHz map since that map has the lowest
angular resolution ($\sim$1.6~arcmin). Next, a linear combination of
the 95 and 150 GHz maps that cancels the tSZ while preserving the
primordial CMB is generated. Lastly, this linear combination map is
added to the 220 GHz map (which is assumed to be tSZ-free since 220
GHz corresponds roughly to the null in the tSZ) with inverse variance
weighting to minimize the noise in the final map.  We note that this
last step, the combination of the 95/150 GHz linear combination data
with the 220 GHz data, could benefit from an optimal weighting of the
two data sets as a function of angular multipole.  The analysis
presented here effectively uses a different, sub-optimal weighting.
We also ignore relativistic corrections to the tSZ spectrum
\citep{itoh98}, which negligibly affect the construction of the
tSZ-free linear combination.

The noise level of the resulting tSZ-free map is roughly 55
$\mu$K-arcmin, significantly higher than the 18 $\mu$K-arcmin noise in
the 150 GHz data: we have sacrificed statistical sensitivity to remove
the tSZ-induced bias.  We use only this tSZ-free linear combination in
the analysis presented here.  Because the kSZ is not frequency
dependent, it is not eliminated with this approach; we will return to
its effects in \S\ref{subsec:ksz}.

\subsection{Galaxy Cluster Catalog}
\label{subsec:catalog}

The galaxy clusters used in this analysis were selected via their tSZ
signatures in the 2500 \degs~SPT-SZ survey as described in
\citet{bleem15b}.  We select all clusters with signal-to-noise $\xi
> 4.5$ and with measured optical redshifts, resulting in 513 clusters.
The clusters analyzed in this work have a median redshift of $z=0.55$
and 95\% of the clusters lie in the $0.14 < z < 1.25$ redshift
interval.  \citet{bleem15b} derived cluster mass estimates for this
sample using a scaling relation between $M_{500}$ and the SZ detection
significance. As described there, the calibration of this scaling
relationship is somewhat sensitive to the assumed cosmology: adopting
the best fit $\Lambda$CDM model from \citet{reichardt13} lowers the
cluster mass estimates by $8\%$ on average, while adopting the best
fit parameters from WMAP9 \citep{hinshaw13} or {\it Planck}
\citep{Planck13-16} increases the cluster mass estimates by $4\%$ and
$17\%$, respectively.  For the cosmological parameters adopted in
\citet{bleem15b} (flat $\Lambda$CDM with $\Omega_m = 0.3$, $h=0.7$,
$\sigma_8=0.8$), the median SZ-derived mass of the cluster sample is
$M_{500} = 3.6\times10^{14}\,\msun$ and 95\% of the clusters lie in
the range $2.5\times10^{14}\,\msun< M_{500} < 9.6\times10^{14}
\,\msun$.  We make use of these mass estimates to generate mock data
in \S\ref{subsec:mock_data} and in \S\ref{sec:spt} we compare these
SZ-derived masses to the cluster masses derived from our measurement
of CMB cluster lensing.

\subsection{Map Cutouts and the Noise Mask}
\label{subsec:masking}

The lensing analysis presented here is performed on ``cutouts'' from
the tSZ-free maps.  Each cutout measures 5.5 arcminutes on a side.
These cutouts are centered on the galaxy clusters' positions
determined in \citet{bleem15b}, and we refer to these as
``on-cluster'' cutouts.
  
For the purposes of null tests (i.e.,~confirming that we observe no
signal when no CMB cluster lensing is occurring), we have produced
many sets of ``off-cluster'' cutouts centered on random positions in
the maps.  To ensure that these off-cluster cutouts have noise
properties representative of the on-cluster cutouts, we draw these
random points from a sub-region of the map that we refer to as the
``noise mask'', defined as follows.  First, for each field we define
the weight map, $w$, which is approximately proportional to the
inverse variance of the instrumental noise at each position in the
map.  Given the weight map of a particular field, we select positions
that have weights between $0.95w_{\rm{min}}$ and $1.05w_{\rm{max}}$,
where $w_{\rm{min}}$ and $w_{\rm{max}}$ are the minimum and maximum
weights at all cluster locations in the field, respectively. Finally,
we exclude from the noise mask any portion of the map that is within
10 arcminutes of an identified point source or cluster.  The point
source catalog used for this purpose is taken from \citet{george14}
and includes all point sources detected at greater than $5\sigma$\,
($\sim $6.4\,mJy at 150 GHz).  For each cluster, we randomly draw 50
off-cluster cutouts from the noise mask region of the field in which
the cluster resides.  This procedure gives us 50 sets of 513
off-cluster cutouts that have the same noise properties as our 513
on-cluster cutouts.  To be robust, our lensing analysis should not
detect any cluster lensing on these off-cluster cutouts, and we
confirm this fact explicitly below.

\section{Analysis}
\label{sec:analysis}

We have developed a maximum likelihood technique for constraining the
CMB cluster lensing signal.  This approach relies on computing the
full pixel-space likelihood of the data given a model for the lensing
deflection angles sourced by a cluster.  The likelihood function
extracts all the information contained in the data about the model
parameters.

The unlensed CMB is known to be very close to a Gaussian random field
\citep[e.g.,][]{planck13-24}.  As such, the likelihood of
observing a particular set of pixelized temperature values, $\vec{d}$,
can be computed given a model for the covariance between these pixels,
$\mathbf{C}$.  The Gaussian likelihood is:
\begin{eqnarray}
\label{eq:gaussian_likelihood}
\mathcal{L}(\mathbf{C}|\vec{d}) = \frac{1}{\sqrt{(2\pi)^{N_{\mathrm{pix}}}\det \mathbf{C}}} \exp \left[ -\frac{1}{2} \vec{d}^T \mathbf{C}^{-1} \vec{d}\right],
\end{eqnarray}
where $N_{\mathrm{pix}}$ is the number of pixels in $\vec{d}$.  Our model for
the data includes contributions from three sources:
\begin{eqnarray}
\mathbf{C} = \mathbf{C}_{\mathrm{CMB}} + \mathbf{C}_{\mathrm{foregrounds}} + \mathbf{C}_{\mathrm{noise}},
\end{eqnarray}
where $\mathbf{C}_{\mathrm{CMB}}$ is the covariance due to the CMB,
$\mathbf{C}_{\mathrm{foregrounds}}$ is the covariance due to signals
on the sky that are not CMB, and $\mathbf{C}_{\mathrm{noise}}$ is the
covariance due to instrumental noise.  In
Eq.~\ref{eq:gaussian_likelihood} we have defined the data vector to be
the deviation from the mean CMB temperature so that $\langle
\vec{d}\rangle = 0$.  

We model the foreground and noise covariances as Gaussian.  The
dominant foreground in our measurement is due to the cosmic infrared
background (CIB).  Although non-Gaussianity is present in the CIB
\citep{crawford14, planck13-30}, the level of non-Gaussianity is
small.  For example, \citet{crawford14} measured the bispectrum of the
220 GHz CIB Poisson term to be $B$$ \sim$1.7$\times 10^{-10} \mu
\rm{K}^3$.  This contributes approximately $B^{2/3}=3.1\times 10^{-7}
\mu \rm{K}^2$ to the power spectrum, which is only $\sim$1\% of the
220 GHz CIB Poisson power spectrum measured by \citet{george14},
$C=4.6\times 10^{-5} \mu \rm{K}^2$.

\subsection{The Lensed CMB Covariance Matrix}

Gravitational lensing is a surface brightness-preserving remapping of
the unlensed CMB.  This means that a photon that is observed at
direction $\hat{n}$ originated from the direction
$\hat{n}_{\mathrm{unlensed}} = \hat{n} + \vec{\delta}(\hat{n})$, where
$\vec{\delta}(\hat{n})$ is the gravitational lensing deflection field.
Lensing thus changes the covariance structure of
$\mathbf{C}_{\mathrm{CMB}}$.\footnote{Our use of a covariance matrix
  (and a Gaussian likelihood) to describe the lensed CMB may result in
  some confusion, as the lensed CMB is known to be non-Gaussian.  The
  lensed CMB is a remapping of a Gaussian random field; by effectively
  undoing this remapping, our likelihood tranforms the observed CMB
  back into a Gaussian random field.  This is possible because we
  construct an explicit model for the lensing deflection field.
  Lensing by LSS complicates this simple picture somewhat because we
  do not construct an explicit model for the deflections sourced by
  LSS.} Since the cluster position is uncorrelated with the CMB
temperature, the mean of the data will remain zero. In principle,
$\mathbf{C}_{\mathrm{foregrounds}}$ can also change as a result of
gravitational lensing if, for instance, some of the foreground
emission is sourced from behind the cluster.  This issue warrants
careful consideration and we will return to it in more detail below.
$\mathbf{C}_{\mathrm{noise}}$ is, of course, unaffected by
gravitational lensing since it is not cosmological.

Because we are interested in the behavior of the CMB on small angular
scales comparable to the sizes of galaxy clusters, a flat sky
approximation is appropriate here and we can replace $\hat{n}$ with
the planar $\vec{x}$.  The calculation of the lensed CMB covariance
matrix, $\mathbf{C}_{\mathrm{CMB}}(M)$, for a cluster of mass $M$ then
proceeds exactly as in the unlensed case
\citep[e.g.,][]{Dodelson:ModernCosmology}, except $\vec{x}$
must be replaced with $\vec{x}_{\mathrm{unlensed}} = \vec{x} +
\vec{\delta}^M(\hat{n})$ (the superscript $M$ here is used to indicate
that the deflection field is a function of the cluster mass).  We find
that the elements of the lensed covariance matrix can be written as
\begin{eqnarray}
\label{eq:lensed_cov_element}
&& \mathbf{C}_{\rm{CMB},ij}(M) = \nonumber \\
&& \int d^2 x \int d^2 x' \, B_i(\vec{x})
B_j(\vec{x}') g(\vec{x} + \vec{\delta}^M(\vec{x}), \vec{x}' +
\vec{\delta}^M(\vec{x}')),\nonumber \\
\end{eqnarray}
where 
\begin{eqnarray}
&& g(\vec{x} + \vec{\delta}^M(\vec{x}), \vec{x}' + \vec{\delta}^M(\vec{x}')) \nonumber \\
&\approx& \sum_{l} C_l \frac{(2l+1)}{4\pi} J_0\left(l \left| (\vec{x}
+ \vec{\delta}^M(\vec{x})) - (\vec{x}' + \vec{\delta}^M(\vec{x}')) \right|
\right),\nonumber \\
\end{eqnarray}
and $J_0$ is the zeroth order Bessel function of the first kind. Here,
$B_i(\vec{x})$ is the pixelized beam and transfer function for pixel
$i$; i.e.,~given a true sky signal $f(\vec{x})$, a noiseless
experiment would measure a signal in pixel $i$ equal to $s_i = \int
d^2x B_i(\vec{x})f(\vec{x})$.  For ease of notation, we lump the
telescope beam and transfer functions into a single object; in
reality, these two functions are sourced by very different mechanisms
as was discussed in \S\ref{sec:data}.  $C_l$ is the power spectrum of
the CMB, which we obtain from CAMB\footnote{http://camb.info}
\citep{lewis99, Howlett:2012} using the best-fit WMAP7+SPT cosmology
from \citet{story13}.  Here we use the lensed CMB power spectrum to
account for the LSS present at redshifts below and above the cluster
redshift.\footnote{By using the LSS-lensed $C_l$'s to compute the
model covariance matrix, we have implicitly assumed that the LSS
lenses the CMB before it is lensed by the cluster.  This approximation
is not completely correct since some structure is presumably located
between us and the cluster.  However, at most, the error introduced by
this approximation could be as large as the product of the
cluster-lensing and the LSS-lensing changes to the covariance matrix
and is therefore very small.  In the absence of a cluster or for a
cluster at $z=0$, our model recovers the exact covariance matrix.}

\subsection{The Deflection Angle Template}

The lensed CMB covariance matrix can be computed from
Eq.~\ref{eq:lensed_cov_element} given a model for the deflection field
sourced by the cluster.  The deflection field can in turn be computed
from a model for the cluster mass distribution if the cluster redshift
is known. In this analysis, we assume a Navarro-Frenk-White (NFW)
profile for the cluster mass distribution, parameterized in terms of
$M_{200}$ and the concentration, $c$ \citep{navarro96}. Written in
this way, the NFW profile is
\begin{eqnarray}
\label{eq:nfw_profile}
\rho(r) = \frac{(200/3) c^3}{\ln (1+c)- \tfrac{c}{1+c}} \frac{\rho_{\mathrm{crit}}(z)}{\left(\tfrac{rc}{r_{200}}\right)\left(1+ \tfrac{rc}{r_{200}}\right)^2},
\end{eqnarray}
where $\rho(r)$ is the mass density a distance $r$ from the center of
the cluster; $\rho_{\mathrm{crit}}(z) = 3 H^2(z) /(8\pi G)$ is the
critical density for closure of the Universe at redshift $z$; and
$r_{200}$ is defined to be the radius at which the mean enclosed
density is $200\rho_{\mathrm{crit}}(z)$.  The mass enclosed within
this radius is $M_{200} = (800\pi/3) \rho_{\mathrm{crit}}(z)
r_{200}^3$. Henceforth, when referring to the cluster mass we will use
$M_{200}$ rather than the more generic $M$.  The concentration
parameter, $c$, controls how centrally concentrated the density
profile is, with higher values of $c$ resulting in a more centrally
peaked mass distribution. Simulations suggest that $c$ is a slowly
varying function of the cluster mass and redshift; for a $M_{200} =
5\times 10^{14}\,\msun$ cluster, the expected concentration is $c\sim
$2.7 \citep{duffy08}.  Since we are concerned with halos of mass
$M_{200} \sim 5 \times 10^{14}\,\msun$ here and because our likelihood
constraints are only weakly sensitive to the concentration, we fix $c
= 3$ throughout.  The results obtained by varying $c$ from 2 to 5 are
essentially identical, as we discuss in
\S\ref{subsubsec:sys_mass_profile}.

While the NFW profile is a common choice for parameterizing the
density profiles of galaxy clusters, true cluster density profiles may
exhibit significant deviations from this form.  High resolution dark
matter-only simulations, for instance, suggest that the density
profiles of the inner cores of clusters are flatter than predicted by
the NFW formula (which diverges as $r^{-1}$ for small $r$)
\citep[e.g.,][]{Merritt:2006,Navarro:2010}.  The introduction of
baryonic effects into such simulations has also been shown to
significantly impact the cluster density profile at small $r$, causing
departures from the NFW form \citep[e.g.,][]{Gnedin:2004, Duffy:2010,
  Gnedin:2011, Schaller:2014}.  Simulations also suggest that for
massive or rapidly accreting halos, the outer density profile ($r
\gtrsim 0.5\,r_{200}$) declines more rapidly than predicted by the NFW
formula \citep[e.g.,][]{Diemer:2014}.  Finally, halos of galaxy
clusters are not expected to be perfectly spherical, but rather
triaxial \citep[e.g.,][]{Jing:2002}.  Still, despite these caveats,
the NFW profile has proven an excellent fit to weak lensing
observations of galaxy clusters.  Although the density profile of an
individual galaxy cluster may exhibit significant deviations from the
NFW form, the profile averaged over many clusters -- such as the 513
clusters considered here -- has been shown to be very well described
by an NFW mass distribution \citep[e.g.,][]{johnston07, Okabe:2010,
  Newman:2013}.  Furthermore, departures from the NFW profile in the
central part of the cluster are unlikely to have much effect on our
results because of the low resolution (roughly 1 arcminute) of our
data, and because the mass of the core is a small fraction of the
total cluster mass.  Ultimately, the NFW profile is more than adequate
for our purposes since the current data set does not have the
resolution or sensitivity to distinguish between different profiles.
We constrain the potential systematic effects introduced into our
analysis by departures from the NFW profile in
\S\ref{subsec:systematics_results}.

For a NFW profile, the deflection vector at angular position
$\vec{\theta}$ away from the cluster is
\begin{eqnarray}
\label{eq:nfw_deflection}
\vec{\delta}^M \left( \theta\right) = -\frac{16 \pi G A}{cr_{200}} \frac{\vec{\theta}}{\theta} \frac{d_{\rm{SL}}}{d_{\rm{S}}} f (d_{\rm{L}} \theta c /r_{200}),
\end{eqnarray}
where $d_{\rm{L}}$, $d_{\rm{S}}$ and $d_{\rm{SL}}$ are the angular
diameter distances to the lens, to the source, and between the source
and the lens, respectively, and $\theta = |\vec{\theta}|$
\citep{bartelmann96,Dodelson:2004}.  The function $f(x)$ is given by
\begin{equation}
f(x) = \frac{1}{x} \begin{cases}
 \ln(x/2) + \frac{\ln\left( x/[1 - \sqrt{1-x^2}]\right)}{\sqrt{1-x^2}}, & \text{if}\ x<1 \\
 \ln(x/2) + \frac{\pi/2 - \arcsin(1/x)}{\sqrt{x^2 - 1}}, & \text{if}\ x>1
\end{cases}
\end{equation}
and the constant $A$ is related to $M_{200}$ and $c$ via
\begin{eqnarray}
A = \frac{M_{200}c^2}{4\pi\left[\ln(1+c) - c/(1+c) \right]}.
\end{eqnarray}
In our analysis we allow the cluster mass to be negative; a negative
cluster mass simply means that the deflection vector is pointed in the
opposite direction of that predicted for a positive cluster mass of
equal magnitude.

\subsection{Numerical Implementation}
\label{subsec:numerics}

With the measured beam and transfer functions of SPT and the
deflection angle template of Eq.~\ref{eq:nfw_deflection}, the
predicted $\mathbf{C}_{\mathrm{CMB}}(M_{200})$ can be computed by
direct integration of Eq.~\ref{eq:lensed_cov_element}.  Unfortunately,
evaluating the 4D integral in Eq.~\ref{eq:lensed_cov_element} is
computationally expensive and the full covariance matrix must be
computed many times.  Consequently, we instead rely on Monte Carlo
simulations to calculate the lensed CMB covariance matrix.

The unlensed covariance matrix is first computed at 1.0 arcminute
resolution across an angular window 70.5 arcminutes on a side (this
wide range relative to the cluster cutouts -- which are only 5.5
arcminute on a side -- ensures that we capture the full effects of the
SPT beam and transfer function).  In the absence of lensing,
Eq.~\ref{eq:lensed_cov_element} can be simplified significantly, and
the unlensed covariance elements can be quickly calculated
\citep[e.g.,][]{Dodelson:ModernCosmology}.  Many Gaussian realizations
of this unlensed covariance matrix (i.e.,~realizations of the unlensed
CMB) are then generated.  Next, a high resolution (0.1 arcminute) map
of the deflection field is generated for a particular $M_{200}$ and
$z$.  The unlensed CMB maps are then interpolated at the positions of
the deflected high-resolution pixels.  Since the primordial CMB is
smooth on scales below a few arcminutes this interpolation is very
accurate.  The resultant maps are then degraded to the resolution of
the tabulated beam and transfer functions, which are applied to the
mock maps using Fast Fourier Transforms.  Finally, the mean of the
product of the lensed temperatures in pairs of pixels, $d_id_j$, is
computed across the many simulated realizations of the lensed
CMB. This mean serves as our estimate of ${\bf
  C}_{\mathrm{CMB}}(M_{200})$.

Our baseline analysis uses 20,000 simulated realizations of the lensed
CMB to form an estimate of the lensed CMB covariance matrix.  To
ensure that this procedure has
reached the precision required for our analysis, we repeat the covariance estimation using fewer and lower-resolution simulations.  We find that 
decreasing the number of simulations by a factor of two, increasing the
pixel size at which the lensing operation is performed by a factor of
2.5, and decreasing the window size from 70.5 arcmin to 60.5
arcmin all lead to small changes in the estimated covariances
matrices (on the order of a few percent).  We also repeat the full likelihood analysis using the degraded covariance estimates and find that the change in the likelihood is entirely negligible (less than a percent in most cases).  We are
therefore confident that our covariance estimation procedure has
acheived sufficient precision for the analysis presented here.

Even when performed in the Monte Carlo fashion described above, the
computation of the lensed CMB covariance matrix is still
computationally expensive.  To speed up the analysis of the data even
more, we compute the lensed covariance matrix across a grid of
$M_{200}$ and $z$; the lensed covariance matrix at the desired mass
and redshift can then be computed via interpolation.  Our baseline
analysis uses 31 evenly spaced $M_{200}$ values and 7 evenly spaced
$z$ values.  To determine whether the accuracy of the covariance
interpolation is sufficient for our measurement, we have increased the
resolution of the $M_{200}$ and $z$ grid across which the covariance
matrix is evaluated and have found the impact on our likelihood
results to be negligible. 

\subsection{Noise and Foreground Covariance}
\label{subsec:noise_and_foreground}

To compute the likelihood in Eq.~\ref{eq:gaussian_likelihood} we must
also estimate $\mathbf{C}_{\mathrm{nf}} \equiv
\mathbf{C}_{\mathrm{noise}} + \mathbf{C}_{\mathrm{foregrounds}}$.  We
take the approach of computing this combination of covariances
directly from the data.  Since the noise level varies somewhat from
field to field, the estimation of $\mathbf{C}_{\mathrm{nf}}$ must be
performed separately for each field.  To do this, we randomly sample
cutouts from the SPT maps of each field to measure the covariance of
the observed data, $\mathbf{C}_{\mathrm{obs}}$.  These samples are
drawn from the noise mask region defined in \S\ref{subsec:masking}.
$\mathbf{C}_{\mathrm{nf}}$ is then estimated by subtracting the
predicted CMB-only covariance from the measured CMB+noise+foreground
covariance, i.e.,~$\mathbf{C}_{\mathrm{nf}} =
\mathbf{C}_{\mathrm{obs}} - \mathbf{C}_{\mathrm{CMB}}(M_{200} = 0)$.

If the foregrounds are lensed by the cluster it is possible for
$\mathbf{C}_{\mathrm{foregrounds}}$ to vary with $M_{200}$.  Modeling
foreground lensing, however, would require knowledge of the redshift
distribution of the foregrounds; for the sake of simplicity we assume
that the foregrounds remain unlensed in our analysis.  We quantify the
bias introduced into our analysis by this assumption using mock data,
as described in \S\ref{subsec:systematics_results}.  For the purposes
of generating this mock data, it is useful to have estimates of both
$\mathbf{C}_{\mathrm{noise}}$ and $\mathbf{C}_{\mathrm{foregrounds}}$
(rather than only the sum $\mathbf{C}_{\mathrm{noise}} +
\mathbf{C}_{\mathrm{foregrounds}}$).  To estimate
$\mathbf{C}_{\mathrm{noise}}$ we sample cutouts from the $L-R$
difference maps described in \S\ref{subsec:cmbdata}.  This sampling
procedure is done using the same noise masks as above so that
$\mathbf{C}_{\mathrm{noise}}$ accurately reflects the noise at the
cluster locations.

$\mathbf{C}_{\mathrm{foregrounds}}$, on the other hand, is estimated
using previous constraints on the power spectra of the dominant
foreground sources.  For the tSZ-free maps that we use in this
analysis, the dominant foregrounds are the `Poisson' and `clustered'
components constrained in \citet{reichardt12b}.  The Poisson
foreground results from point sources below the detection threshold
that are randomly distributed on the sky and has $C_l = C_0$,
independent of $l$.  The amplitude of the Poisson component is
estimated from the data.  The clustered foreground model accounts for
the clustering of point sources and is modeled as $D_l \equiv C_l
l(l+1)/(2\pi) = D_0$ independent of $l$ for $l < 1500$, and $D_l
\propto l^{0.8}$ for $l > 1500$.  The amplitude of the clustered
component is taken from \citet{reichardt12b}, adjusted to account for
the fact that our maps are constructed from a weighted combination of
observations at three frequencies.  With the foreground power spectra
determined, $\mathbf{C}_{\mathrm{foregrounds}}$ can be calculated in
the same way as the unlensed CMB covariance matrix.

We emphasize that the main analysis estimates
$\mathbf{C}_{\mathrm{nf}} \equiv \mathbf{C}_{\mathrm{noise}} +
\mathbf{C}_{\mathrm{foregrounds}}$ directly from the data, and that
the individual estimates of $\mathbf{C}_{\mathrm{noise}}$ and
$\mathbf{C}_{\mathrm{foregrounds}}$ are used only to test for certain
systematic effects using mock data.

\subsection{Combining the Likelihoods}
\label{subsec:combining_likelihoods}

With our estimates of $\mathbf{C}_{\mathrm{CMB}}(M_{200})$ and
$\mathbf{C}_{\mathrm{noise}} + \mathbf{C}_{\mathrm{foregrounds}}$, we
now have all the ingredients necessary to evaluate the likelihood in
Eq.~\ref{eq:gaussian_likelihood}.  For a cutout around the $i$th
cluster, we evaluate the likelihood, $\mathcal{L}_i(M_{200})$, as a
function of $M_{200}$ to constrain the effects of CMB lensing by that
cluster.  However, since the instrumental noise is large relative to
the CMB cluster lensing signal, we do not expect to obtain a detection
of the lensing effect around a single cluster.  Instead, we must
combine constraints from multiple clusters.  One way to accomplish
this is to compute the likelihood
$\mathcal{L}_{\mathrm{total}}(M_{200}) =
\prod_i^{N_{\mathrm{clusters}}} \mathcal{L}_i(M_{200})$, where
$N_{\mathrm{clusters}} = 513$ is the number of clusters in our sample.
This method of combining likelihoods is appealing because it is simple
and because it depends only on the lensing information.

Not all the masses in the sample are the same, so the above treatment
-- which assumes all clusters share a common mass -- provides more of
an estimate of the detection significance than any useful information
on the masses of the clusters in the sample. Furthermore, the spread
in masses will likely lead to a spread in the width of the likelihood
function, i.e.,~a degradation in the signal-to-noise. Some of this can
be recaptured by scaling the $M_{200}$ parameter for each cluster by
an external mass estimator for that cluster, and indeed estimates of
each cluster's mass can be obtained from the strength of the SZ signal
at the cluster location.  Here we use the SZ-determined cluster masses
from \citet{bleem15b} that were discussed in \S\ref{subsec:catalog}.
We convert the $M_{500,\mathrm{SZ}}$ measured in \citet{bleem15b}
into $M_{200,\mathrm{SZ}}$ using the \citet{duffy08}
mass-concentration relation.  So an improved likelihood that includes
this information is written not as a function of $M_{200}$, but rather
as
\begin{equation}
\mathcal{L}_i \rightarrow \mathcal{L}_i\left(\frac{M_{200}}{M_{200,\mathrm{SZ}}}\, M_{200,\mathrm{SZ},i}\right),
\label{eq:scale}
\end{equation}
with a new free global parameter $M_{200}/M_{200,\mathrm{SZ}}$.  The
individual cluster likelihoods expressed as functions of
$M_{200}/M_{200,\mathrm{SZ}}$ can then be combined as before:
\begin{eqnarray}
\mathcal{L}_{\mathrm{total}}\left(\frac{M_{200}}{M_{200,\mathrm{SZ}} }\right) =
\prod_i^{N_{\mathrm{clusters}}} \mathcal{L}_i \left(\frac{M_{200}}{M_{200,\mathrm{SZ}}}\, M_{200,\mathrm{SZ},i}\right). \nonumber \\
\end{eqnarray}
Note, however, that
any intrinsic scatter in the relationship between the lensing-derived
$M_{200}$ and the SZ-derived $M_{200,\mathrm{SZ}}$ will lead to
additional broadening of the combined multi-cluster likelihood as a
function of $M_{200}/M_{200,\mathrm{SZ}}$.  We will employ both
methods of combining individual cluster likelihoods in
\S\ref{sec:spt}.

\begin{figure*}[ht]
\center
\includegraphics[scale = 0.8]{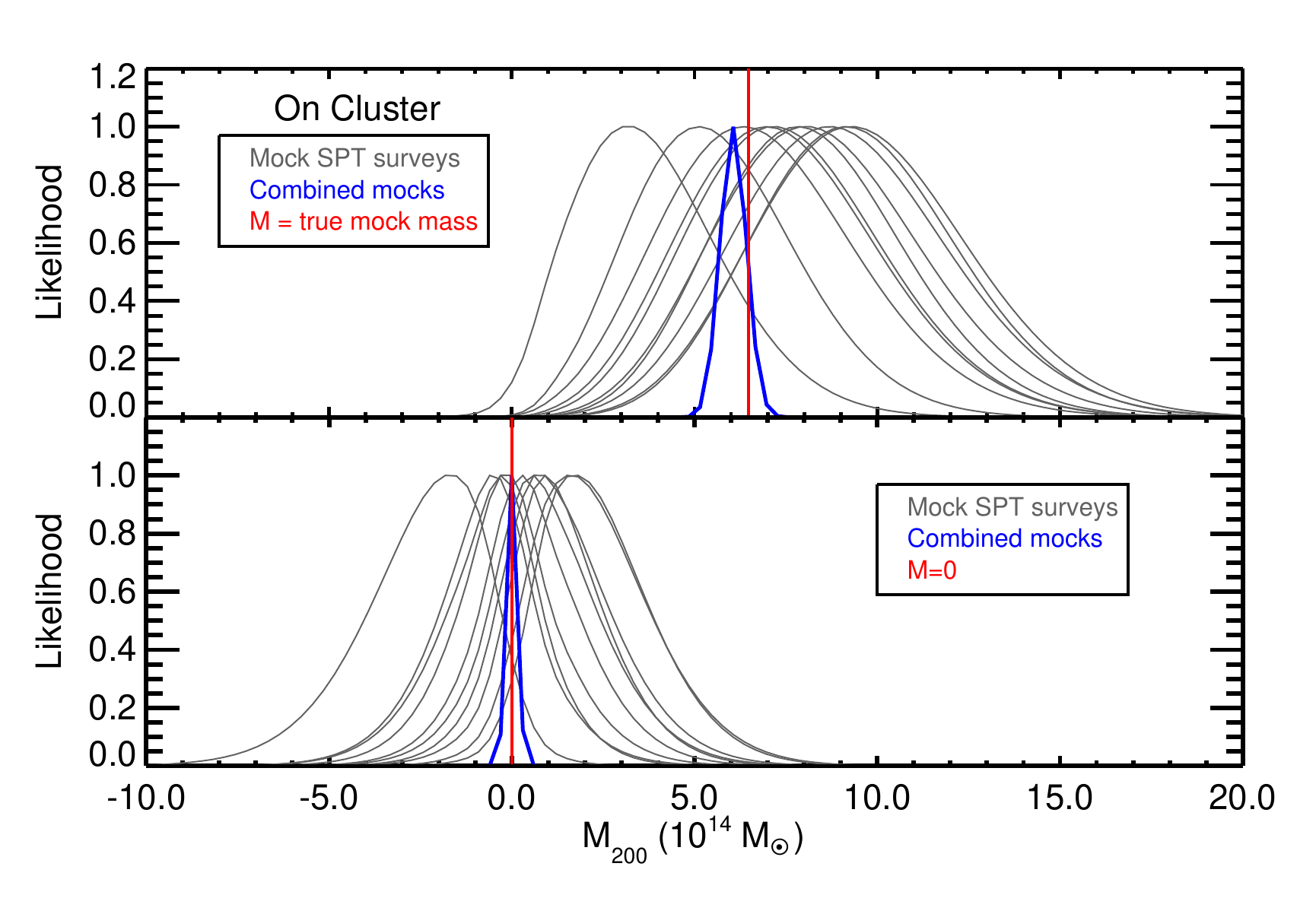}
\caption{Constraints on $M_{200}$ from the analysis of mock data that
  is designed to mimic real data from the SPT. The top panel shows the
  likelihood as a function of $M_{200}$ for patches centered on
  clusters; the bottom panel shows the same for patches centered at
  random points ({\it off-cluster}). Each gray curve represents the
  constraint obtained from a single realizations of an SPT-like survey
  that detects 513 clusters; the blue curves are combined constraints
  from 50 such realizations. }
\label{fig:likelihood_results_mocks}
\end{figure*}

\subsection{Mock Data}
\label{subsec:mock_data}

In order to test our analysis pipeline and study possible sources of
systematic error we generate and analyze mock data.  The mock data
sets include contributions from the lensed (and unlensed) CMB,
foregrounds and noise.  The mock cluster redshift distribution is
identical to the redshift distribution of the real clusters.  To
generate cluster masses for our mock catalog, we convert the
SZ-derived $M_{500}$ values described in \S\ref{subsec:catalog} to
$M_{200}$ assuming that the clusters are described by NFW profiles
with the \citet{duffy08} mass-concentration relation.  The resultant
sample has a median mass of $M_{200} = 5.6\times10^{14}\,\msun$ and
95\% of the clusters have $4.0\times10^{14}\,\msun < M_{200} <
1.37\times10^{15}\,\msun$.

For each mock cluster, a realization of the lensed and unlensed CMB
was generated in the same manner described in \S\ref{subsec:numerics}.
The clusters were distributed among the SPT fields identically to the
real clusters, and the appropriate beam and transfer functions for
each field were applied.  Gaussian realizations of the measured noise
and foreground covariance matrix, $\mathbf{C}_{\mathrm{nf}}$, were
added to the mock data in a field-dependent fashion.  The process of
generating a mock cluster catalog was repeated 50 times to build
statistics.  Each mock catalog includes entirely new realizations of
the CMB, foregrounds and noise.

\section{Results on Mock Catalogs}
\label{sec:mock_results}

\subsection{Projections}
\label{subsec:projections}

The results of our analysis of the mock cluster cutouts are shown in
Fig.~\ref{fig:likelihood_results_mocks}.  The top panel shows the
results of analyzing the mock data when CMB lensing is turned on,
while the bottom panel shows the results when CMB lensing is turned
off (i.e.,~a null test).  Each gray curve represents the combined
likelihood constraints from an SPT-like survey with 513 clusters
generated in the manner described above; the blue curves show the
combined constraints from 50 mock data sets of 513 clusters.  The
vertical red line in the top panel indicates the true mean cluster
mass in the mock survey. Each mock data set strongly prefers a
positive cluster mass over $M_{200} \leq 0$. The combined constraint
from 50 mock data sets in Fig.~\ref{fig:likelihood_results_mocks}
illustrates that the likelihood prefers the mean cluster mass of the
sample.  When the analysis is performed on the unlensed mock data
(bottom panel), none of the 50 mock data sets yield a significant
detection, and the mean is centered at the (correct) value of $M_{200}
= 0$.

To quantify the significance of our measurement of CMB cluster lensing
(for both mock and real data) we use a likelihood ratio test.  Since
we are interested in whether or not the data prefer lensing over the
null hypothesis of no lensing (i.e.,~$M_{200} = 0$), we define the
likelihood ratio
\begin{eqnarray}
\Lambda = \frac{\mathcal{L}(M_{200} = 0)}{\max\mathcal{L}(M_{200})}.
\end{eqnarray}
In the large sample size limit (i.e., many clusters), $-2\ln \Lambda$
should be $\chi^2(k=1)$-distributed with $k=1$ degree of freedom.
Note that this statement does not assume that the likelihood for each
cluster is Gaussian as a function of $M_{200}$.  The $p$-value for the
measurement is then found by integrating the $\chi^2(k=1)$
distribution below $-2\ln \Lambda$.  Our reported detection
significance is calculated by converting this $p$-value into a
standard, two-sided Gaussian significance and is exactly equal to
$\sqrt{-2\ln \Lambda}$.  All detection significances are reported in
this way below.  Averaging across the 50 mocks discussed above, we
find that the mean detection significance for an SPT-like survey
(i.e.,~513 mock clusters) is $3.4 \sigma$.

\subsection{Systematics Tests}
\label{subsec:systematics_results}

Several sources of systematic error can potentially affect our CMB
cluster lensing measurement.  We quantify the impact of these
systematic effects on our analysis by modeling them in mock data.  For
the purposes of these systematic tests we generate new mock data
consisting of 500 realizations of the CMB, noise, and foregrounds for
a single cluster with $z = 0.55$ and $M_{200} =
5.6\times10^{14}\,\msun$, corresponding to the median redshift and
SZ-derived mass for clusters in our sample.  Various systematic
effects are introduced to this mock data set as described below.  We
then analyze the mock data {\it neglecting} the presence of the
systematic effects and measure how the likelihood changes.

We express the bias introduced by each systematic as the fractional
shift in the maximum likelihood mass: $(M^{\rm{ML}}_{\rm{sys}} -
M^{\rm{ML}})/M^{\rm{true}}$, where $M^{\rm{ML}}_{\rm{sys}}$ is the
maximum likelihood mass in the presence of the systematic,
$M^{\rm{ML}}$ is the maximum likelihood mass without the systematic,
and $M^{\rm{true}} = 5.6\times10^{14}\,\msun$ is the true mass of the
mock clusters.  This process is repeated 50 times and we report the
mean value of the bias across these trials.  We caution that this
procedure is not meant to rigorously quantify the systematic error
budget of our lensing constraints; we have, after all, assumed a
single mass and redshift for all of the mock clusters.  Instead, these
estimates are provided for two purposes.  First, they suggest that the
individual systematic errors associated with our cluster mass
measurement are likely small compared to the statistical error bars on
this measurement.  Second, the estimates provided below highlight the
relative importance of each of the systematic effects that we consider
here.

\subsubsection{Monopole Contamination}

The first systematic that we consider is anything that leads to a
signal at the cluster center (a ``monopole'').  The CMB cluster
lensing signal vanishes at the cluster center and therefore has no
monopole component.  Since our model includes no other signals
correlated with the cluster, any residual monopole-like signal at the
cluster location is not included in our model and could therefore bias
our analysis.  One important potential source of monopole
contamination is residual tSZ in our tSZ-free maps.  Although the
linear combination map used is nominally independent of tSZ, the
finite width of the observing bands and relativistic corrections to
the tSZ \citep{itoh98} can produce a small residual component. Other
potential sources of monopole contamination include the integrated
dusty emission or radio emission from cluster member galaxies much too
faint to be individually detected in SPT maps. Strong emission from
individual cluster members is treated in the next section.

We determine the amplitude of such contamination directly from our
data.  Stacking all of the cluster cutouts reveals that the level of
monopole contamination is consistent with a $\beta$ profile
\citep{cavaliere76,cavaliere78} with $\beta = 1$, $\theta_c = 0.5$
arcmin, and an amplitude of $-3$ \muK\, for each cluster.  We
introduce this level of contamination into our 50 sets of 500 mock
cutouts and repeat the likelihood analysis (just as before, without
accounting for the monopole contamination) to determine how our
likelihood constraints are affected.  Across 50 sets of mock cutouts,
we find that monopole contamination of the measured amplitude leads to
a shift in the maximum likelihood mass that is $\lesssim 1\%$, well
below the statistical precision of our cluster mass constraint.

\subsection{Emission from Individual Cluster Members}

The contamination of our measurement by a single bright cluster galaxy
does not in general behave like the monopole contamination considered
above. In particular, a single source could fill in one side of the
cluster lensing dipole if its projected position relative to the
cluster is at a particular radius and orientation.  At 150 GHz and a
resolution of 1.6 arcmin, a 1 mJy source will have an equivalent CMB
fluctuation temperature of $10$\muK\, and, assuming a spectral index
of $\alpha = -0.5$, will have a temperature fluctuation of roughly
$-$10\muK\, in our tSZ-free maps.  We simulate the effects of such
sources on our analysis by introducing a single point source with
beam-smoothed amplitude of $-$10\muK\, into each of our mock cutouts.
We choose the location of the point source randomly across a disk of
radius 1.5 arcmin centered on the cluster.  Since the CMB cluster
lensing dipole is expected to peak at \,$\sim$1 arcmin away from the
cluster center, sources located much farther than this should have
little effect on our measurement.

We find that introducing this level of point source contamination into
our mock data causes the inferred cluster mass to be biased low by
\,$\sim$7\% on average across our 50 sets of 500 mock cluster cutouts.
In reality, however, not every cluster is expected to have an
associated point source of this magnitude and proximity to the
cluster.  Using the \citet{dezotti10} model for radio source counts at
150 GHz and the results of \citet{coble07}, we estimate that only
\,$\sim$5\% of SPT-SZ clusters will have a 1 mJy or greater source
within 1.5 arcmin of the cluster center.  We only consider radio
sources in this calculation because models of dusty sources predict
fewer bright sources \citep[e.g.,][]{negrello07}, and because star
formation is suppressed in cluster environments
\citep[e.g.,][]{bai07}.  The resulting bias on the mean mass of our
cluster sample would thus be \,$<$1\%, well below our statistical
precision.

\subsubsection{kSZ}
\label{subsec:ksz}

The second systematic that we consider is the kSZ effect.  The kSZ
effect results from scattering of CMB photons with electrons that have
bulk velocities relative to the Hubble flow.  Motions of cluster
electrons could be due, for instance, to the cluster falling towards
nearby superstructures or because the cluster is rotating. While
typically much smaller than the tSZ effect, the kSZ effect is frequency
independent when expressed as a change in brightness temperature, so
the tSZ-free linear combination map contains a kSZ component.

The diffuse kSZ caused by linear or quasi-linear structure will act
only as a source of noise in this analysis, and, because its amplitude
is much smaller than the instrumental noise \citep{george14}, it can
be safely ignored here.  Instead we turn our attention to the kSZ due
to the galaxy clusters themselves.  This cluster kSZ signal will have
two components: a component due to the bulk motion of the cluster, and
a component due to internal velocities. 

To include the effects of the bulk component of the kSZ in our mock
data we rely on the work of \citet{sehgal10}, which used $N$-body
simulations and models for the gas physics at different redshifts to
generate maps of the kSZ effect.  The \citet{sehgal10} kSZ maps are
generated by assigning a single velocity to all gas associated with
each cluster, and thus provide an estimate of the kSZ signal due to
the bulk velocity of each cluster.  The simulated kSZ signal is
introduced into our mock cutouts by extracting cutouts from the
\citet{sehgal10} kSZ maps around clusters with $M_{200}$ between
$5.0\times10^{14} \,\msun$ and $6.0\times10^{14}\, \msun$.  This
selection ensures that the kSZ signal is reasonably well matched to
our mock clusters, which have masses of $5.6\times10^{14}\,\msun$.
The likelihood analysis of the mock cutouts with kSZ is then performed
as before, ignoring the presence of the kSZ.

Across 50 realizations of the mock data, the introduction of a
bulk-velocity kSZ component causes the maximum likelihood mass to be
biased low by $9\%$ on average, below the statistical precision of
this work.  We note that our analysis of mock data with kSZ suggests
that the size of the bias introduced by the presence of the kSZ
depends on the level of instrumental noise and foregrounds in the
data.  If the foreground or instrumental noise contributions are very
small, the bias introduced by the kSZ can become significant.  Future
experiments with higher sensitivity may need to take a more careful
approach to accounting for the kSZ.

The mock kSZ signal considered above does not include the effects of a
kSZ signal due to internal motions of gas within the cluster.  Of
particular concern is the kSZ signal resulting from cluster rotation,
which we call rkSZ.  A cluster that is rotating will induce a
dipole-like kSZ signal since one side of the cluster will be moving
towards us while the other will be moving away.  Consequently, even
though the rkSZ is expected to be small, it is a potentially serious
contaminant for the CMB cluster lensing measurement because of its
similar morphology on the sky.  Unlike the CMB cluster lensing signal,
though, the rkSZ dipole is not preferentially aligned with the
gradient of the CMB temperature field.

Our model for the rkSZ signal is based on the model of
\citet{Chluba:2002}, where it is assumed that a galaxy cluster rotates
as a solid body, motivated in part by the work of \citet{Bullock:2001}
and \citet{Cooray:2002}.  Modeling the electron number density as a
$\beta$-profile, \citet{Chluba:2002} derive an expression for the rkSZ
signal:
\begin{eqnarray}
\label{eq:rksz}
\frac{\Delta T_{\mathrm{rkSZ}}}{T_{\mathrm{CMB}}} (\theta, \phi)
&=& A_{{\rm rkSZ}} \theta \sin i\sin \phi \left( 1 +
\frac{\theta^2}{\theta_{\mathrm{core}}^2} \right)^{1/2 - 3\beta/2},\nonumber \\
\end{eqnarray}
where $A_{{\rm rkSZ}}$ is a parameter that controls the amplitude of
the signal, $\theta$ is the angular distance from the cluster center,
$\phi$ is the transverse angular coordinate and $i$ is the inclination
angle of the cluster.  We set $\beta = 1$ and $\theta_{\mathrm{core}}
= 1$ arcminute as these values are fairly typical for the clusters in
our sample.

The amplitude of the rkSZ signal, $A_{{\rm rkSZ}}$, is not very well
constrained at present.  Simulations
\citep[e.g.,][]{Nagai:2003,Fang:2009,Bianconi:2013} suggest that the
rotational velocities of clusters are typically small compared to the
cluster velocity dispersion.  However, in clusters that have recently
experienced mergers, the rotational velocities may be significantly
larger.  \citet{Chluba:2002} argue that typical peak rkSZ signals are
in the range $0.1$--10\muK, but could be as high as 100\muK\, for a
recent merger.

The model rkSZ signal is introduced into our 50 sets of 500 mock
cutouts assuming a constant value of $A_{{\rm rkSZ}}$ for all mock
clusters.  Each cluster's inclination angle and orientation on the sky
are chosen randomly, however, so the mock rkSZ signal varies from
cluster to cluster.  We explore several values of $A_{{\rm rkSZ}}$,
chosen such that the maximum amplitude of the rkSZ signal (i.e., for
an optimally aligned cluster) varies between 1\muK\, and 20\muK.  We
find that the presence of rkSZ in the mock data acts to reduce our
measured signal. At a maximum amplitude of 1\muK\, the rkSZ introduces
a mass bias of less than $1\%$ to our mass constraints, at 5\muK\, the
peak of the likelihood is biased to lower masses by roughly $8\%$, at
10\muK\, the bias is roughly $28\%$ and at 20\muK\, the bias is $93\%$.
Therefore, it appears that as long as the rkSZ signal is
$\lesssim$10\muK\,, the bias introduced into our mass constraints by
such a signal is less than the statistical precision of this work.
Since most clusters are expected to have rkSZ signals less than
10\muK, we do not attempt to correct for this effect here.  Although
clusters that have experienced recent mergers may have rkSZ signals
that are higher than 10\muK, the number of such clusters in our sample
is likely small.  

\subsubsection{Foreground Lensing}
\label{subsubsec:sys_foreground_lensing}

As discussed above, the degree to which foreground emission is lensed
by the cluster is not very well constrained.  The CIB -- which
constitutes the dominant source of foreground emission -- is thought
to originate from redshifts $z \sim 0.5 \textrm{ to } 4$.  Since our
cluster sample is drawn from $0.05 \lesssim z \lesssim 1.5$, the
amount by which the foregrounds are lensed will likely vary from
cluster to cluster.  Our analysis, however, assumes that foregrounds
remain unlensed.  To investigate the effects of this assumption on our
analysis, we generate mock cutouts with lensed foregrounds assuming
that foreground emission originates from $z=4$. Since the CIB is known
to originate from $z \lesssim 4$, setting $z=4$ gives an approximate
upper bound to the effects of gravitational lensing on the
foregrounds, and therefore an upper limit to the systematic error
introduced into our analysis by assuming no foreground lensing.

Realizations of the lensed clustered foreground can be generated using
the procedure described in \S\ref{subsec:mock_data}.  Lensing the
Poisson foreground is more difficult as this foreground has power
extending to arbitrarily small scales, including scales below that at
which we generate map realizations.  To get around this, we calculate
the lensed Poisson covariance matrix directly from the integral in
Eq.~\ref{eq:lensed_cov_element} and use this covariance matrix to
generate realizations of the lensed Poisson foreground.  The mock
cutouts with lensed foregrounds are then analyzed as before, assuming
that both foregrounds remain {\it unlensed}.

Across the 50 sets of 500 mock clusters that we have generated, we
find that lensing of the foregrounds causes our $M_{200}$ constraint
to be biased low.  Lensing of the Poisson foreground contributes the
dominant part of this bias, owing to its large contribution to the
total covariance relative to that of the clustered foreground.  The
average mass bias introduced into our mock analysis by lensing of the
foregrounds is $7\%$.  We do not correct for this bias, as doing so
would require a detailed modeling of the redshift distribution of the
CIB.  We emphasize, though, that the bias measured here is necessarily
an overestimate of the true bias introduced by foreground lensing
because we have placed the foregrounds at $z=4$ when the true
foreground emission results from $z \leq 4$.

\subsubsection{Cluster Miscentering}
\label{subsubsec:sys_miscentering}

The cluster centers used in our analysis are derived from SPT
measurements of the cluster SZ signal and will generally differ from
the centers of mass of the clusters.  A similar miscentering problem
arises in the context of galaxy shear measurements, where the cluster
center is typically defined as the location of the brightest cluster
galaxy (BCG), even though the BCG may not correspond to the true
center of mass of the cluster \citep[e.g.,][]{vonderlinden14a}.
In that context, cluster miscentering can be a significant source of
systematic error in cluster mass measurements, causing the masses of
miscentered clusters to be underestimated.

We model the effects of imperfect knowledge of the cluster center by
applying random positional shifts to our mock cluster data.  These
offsets are drawn from a two dimensional Gaussian with $\sigma = 30$
arcseconds.  In this model, 68\% of the offsets are smaller than 45
arcseconds.  As a point of reference, \citet{song12b} found that 68\%
of the offsets between SPT-estimated centers and BCGs were smaller
than 38 arcseconds, so the miscentering error introduced here is
likely an overestimate.  Analyzing the miscentered mock data reveals
that the peak likelihood is biased to lower mass by roughly $6\%$ on
average, below the statistical precision of our lensing mass
constraint.  Accurately modeling the size of the miscentering
systematic error would require an understanding of how the
miscentering error varies with cluster mass and redshift, and we do
not attempt such a detailed analysis here.  

\subsubsection{Uncertainty in the Cluster Mass Profile}
\label{subsubsec:sys_mass_profile}

Our analysis assumes a NFW profile for each cluster with concentration
$c=3$.  In reality, the halo concentration is known to vary with
cluster mass and redshift, and to exhibit significant scatter.  To
explore the effects on our analysis of changing the halo
concentration, we regenerate the mock cluster data using halos of
concentration $c=2.5$ and $c=5$.  These two values of the
concentration should bracket the expected range of concentrations
allowed for the clusters in our sample, including effects of
uncertainty in the assumed cosmological parameters
\citep{Dutton:2014}.  The data are then analyzed as before, assuming
$c=3$.  We find that changing the concentration has an essentially
negligible effect on our analysis, which is not surprising given that
our constraints are not sensitive enough to distinguish between
slightly different behaviors of the inner mass profile.  Across 50
realizations of 513 mock clusters, we find that the maximum likelihood
mass increases on average by less than 1\% when $c=5$, well below the
statistical precision of our measurements.  When $c=2.5$ we find that
the maximum likelihod mass decreases by about 1$\%$.  The effects of
changing halo concentration can therefore be safely ignored in this
analysis.

A related source of potential bias is halo triaxiality.  It is well
known from simulations \citep[e.g.][]{Jing:2002,Kasun:2005} that halo
density profiles are not perfectly spherical. Deviations from
sphericity could introduce a bias into our analysis because we have
assumed a perfectly spherical NFW profile.  \citet{Corless:2007} have
found that in the context of traditional galaxy shear measurements,
fitting a spherical NFW profile to the extreme case of a halo
elongated along the line of sight can lead to a 50\% mass bias.
Averaged over all possible halo orientations, however,
\citet{Corless:2007} find that the mean recovered mass is very close
to the true mass.  Given the low sensitivity of our mass constraints
and the findings of \citet{Corless:2007}, it is unlikely that halo
triaxiality has a significant impact on our results.  A detailed
modeling of the effects of halo triaxiality is beyond the scope of
this work.

Finally, we consider deviations of the halo profile from the NFW form
itself.  While large deviations from the NFW profile are expected in
the central region of the dark matter halo, the roughly 1.6 arcminute
resolution of the SZ-free maps means that we are not very sensitive to
the behavior of the density profile in this regime.  Deviations from
the NFW form are also expected for massive clusters in the outskirts
of the halo, $r \gtrsim 0.5 r_{200}$ \citep{Diemer:2014}.  Assuming
the SZ-derived masses described in \S\ref{subsec:catalog}, the median
$\theta_{200} = r_{200}/d_A(z)$ for the clusters in our sample is 5.2
arcminutes, where $d_A(z)$ is the angular diameter distance to the
cluster. This means that our angular window of 5.5 arcminutes around
each cluster is probing $r \sim 0.5 r_{200}$.  Consequently,
deviations from the NFW form in the $r \gtrsim 0.5 r_{200}$ regime
could potentially introduce a systematic error into our mass
constraints.

We model the effects of deviations from the NFW density profile by
approximating the results of \citet{Diemer:2014}.  Mock data with a
non-NFW deflection profile are generated and analyzed assuming the
usual NFW deflection formula. We find that modifying the form of the
deflection profile in this way biases the best fit mass low by roughly
$10\%$ on average across our 50 sets of 500 mock cluster cutouts. 

\subsubsection{Large Scale Structure}

Our NFW lensing template (Eq. \ref{eq:nfw_deflection}) accounts only
for deflections of CMB photons caused by the cluster itself.  It
therefore ignores deflections that could be caused by the presence of
LSS near the line of sight to the cluster.  Lensing by LSS
unassociated with the cluster changes the covariance properties of the
CMB in a well-known way \citep[e.g.][]{seljak96b}.  This effect is
approximated in our model through the use of the LSS-lensed $C_l$'s in
computing the model covariance matrix
(Eq. \ref{eq:lensed_cov_element}).  However, it is well known that
clusters live in overdense environments.  Lensing induced by LSS that
is associated with the cluster is not included in our model and could
therefore bias our analysis.

In the language of the halo model, we have effectively ignored the
two-halo contribution to the lensing signal.  However, weak lensing
data \citep[e.g.][]{johnston07} suggest that within a few virial radii
of the cluster center, the one-halo term dominates the lensing signal.
Since the analysis presented here considers a small angular region
around each cluster that extends to only $\leq 1$ virial radius, it is
safe to neglect the two-halo term in this analysis.

\subsubsection{Cluster Selection}

One remaining potential systematic is related to the SZ-selection
method.  The SPT clusters have been selected at the locations of
decrements in the 95 and 150 GHz maps.  Simulations show that clusters
selected in this fashion will preferentially sit on decrements in the
CMB, and this effect could potentially bias the mass inferred from CMB
lensing.  However, the bias in the background CMB is small, on the
order of -1\muK, and the resulting effect on the CMB lensing mass
is likely to be small compared to our statistical error.

\subsubsection{Combined Systematic Effects}
\label{subsubsec:combined_sys}

The above discussion has considered how several different systematic
effects can individually bias our lensing constraints.  We now attempt
to estimate the total bias resulting from the combination of multiple
systematic effects.  Our combined systematic model includes the five
most significant biases considered above.  We include the bulk motion
kSZ, the rkSZ with peak amplitude of 5\muK, and foreground lensing as
described in \S\ref{subsubsec:sys_foreground_lensing}.  The clusters
are miscentered as described in \S\ref{subsubsec:sys_miscentering} and
the cluster density profile used is the \citet{Diemer:2014} profile
described in \S\ref{subsubsec:sys_mass_profile}.  We find that the
mean bias introduced by this combined systematic model is a 39\% bias
to lower cluster mass.  The measured bias is consistent with the
product of the individual biases (34\%), given the scatter among the
50 simulation realizations.  A 39\% bias to lower cluster mass amounts
to a roughly $0.85\sigma$ shift in units of the statistical
uncertainty.  
This should be interpreted as an approximate upper limit on the 
bias to lower cluster mass, as we have placed all of the foreground 
emission at $z=4$ and have likely over-estimated the effect of miscentering.
We do not attempt to correct for this systematic bias, although doing so would not alter the main 
conclusions of this work, as discussed in Section~\ref{sec:conclusion}.

\begin{figure*}
\center
\includegraphics[scale = 0.8]{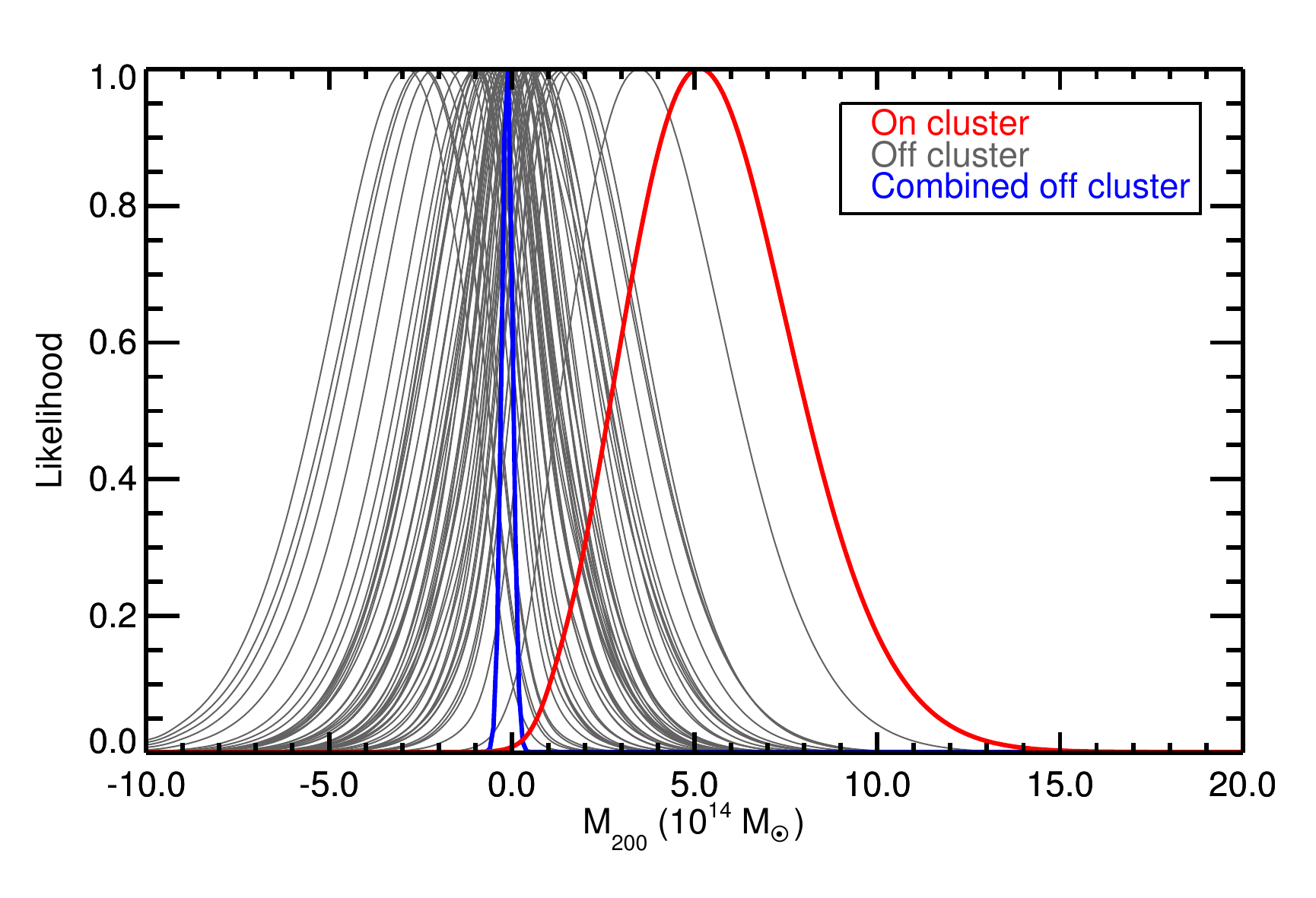}
\caption{Constraints on $M_{200}$ resulting from analysis of SPT
  data. The thick red curve is our result for 513 on-cluster cutouts,
  while each thin gray curve is our result for a separate realization
  of 513 off-cluster cutouts.  The combined constraint from the 50
  sets of off-cluster cutouts is the thick blue curve.}
\label{fig:likelihood_results}
\end{figure*}

\begin{figure*}
\center
\includegraphics[scale = 0.8]{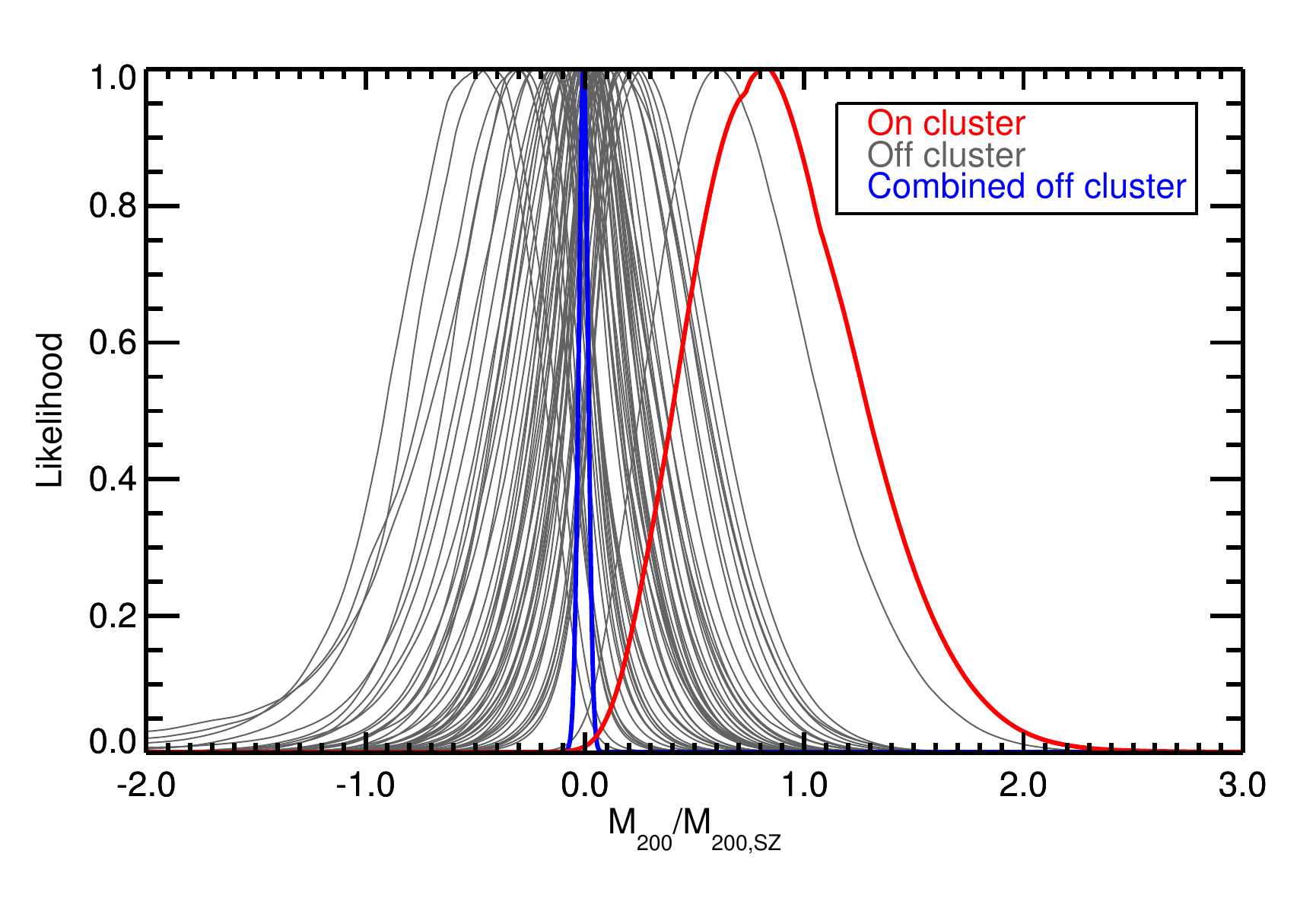}
\caption{Same as Fig.~\ref{fig:likelihood_results} except the
  likelihood has been computed as a function of
  $M_{200}/M_{200,\mathrm{SZ}}$, where $M_{200,\mathrm{SZ}}$ is the
  cluster mass computed from the measured tSZ signal as described in
  \citet{bleem15b}.}
\label{fig:likelihood_results_szscaled}
\end{figure*}

\section{Results}
\label{sec:spt}

Fig.~\ref{fig:likelihood_results} shows the results of our likelihood
analysis applied to the data described in \S\ref{sec:data}.  The red
curve represents our constraint from the analysis of 513 on-cluster
cutouts.  Each of the 50 gray curves represents the constraint
from 513 off-cluster cutouts chosen from the same fields as the
on-cluster cutouts in the manner described in \S\ref{sec:data}.  The
thick blue curve is the combined constraint from the 50 sets of
off-cluster cutouts.

The on-cluster likelihood in Fig.~\ref{fig:likelihood_results} shows a
preference for positive mass.  We find that $M_{200} = 0$ is ruled out
at $3.1\sigma$ using the likelihood ratio test described above.  We
assume a flat prior on $M_{200}$ so that the posterior probability of
$M_{200}$ is directly proportional to the likelihood.  Integrating the
posterior on $M_{200}$ yields a 68\% confidence band of $M_{200} =
5.1_{-2.1}^{+2.5} \times 10^{14} \,\msun$.  The results of our
analysis of the SPT clusters are also consistent with the projections
from mock data, which had a mean detection significance of
3.4$\sigma$.  The off-cluster likelihoods shown in
Fig.~\ref{fig:likelihood_results} (gray curves) are consistent with
$M_{200} = 0$.  The worst null likelihood has a detection significance
of $2.2\sigma$, which is reasonable since we have considered 50 null
likelihoods.  The combined constraint from all 50 null likelihoods is
also consistent with $M_{200} = 0$ at $0.84 \sigma_{\rm 50}$, where
$\sigma_{50}$ is the standard deviation computed from the 50 stacked
likelihoods.  There is therefore no evidence of any bias in our
off-cluster analysis.

As described in \S\ref{subsec:combining_likelihoods}, the constraints
on the lensing mass $M_{200}$ of our cluster sample can be translated
into constraints on the ratio between the lensing mass,
$M_{200,\mathrm{lens}}$, and the cluster mass estimated from the tSZ
effect, $M_{200,\mathrm{SZ}}$.  The likelihood curve of the ratio
$M_{200,\mathrm{lens}}/M_{200,\mathrm{SZ}}$ is calculated per cluster,
and the combined constraint (assuming a flat prior on the ratio) is
\begin{equation}
\frac{M_{200,\mathrm{lens}}}{M_{200,\mathrm{SZ}}} = 0.83_{-0.37}^{+0.38}\qquad ({\rm 68\%} \,C.L.) 
\end{equation}  
The mean mass inferred from CMB cluster lensing is consistent with the
mean mass inferred from the tSZ signal at $0.5\sigma$.  This
constraint and the corresponding off-cluster likelihoods are shown in
Fig.~\ref{fig:likelihood_results_szscaled}.  Using the likelihood
ratio test described above, we find that
$M_{200,\mathrm{lens}}/M_{200,\mathrm{SZ}} = 0$ is ruled out at
$3.1\sigma$.

As pointed out in \S\ref{subsec:catalog}, depending on the assumed
cosmological model, the mean SZ-derived cluster mass can vary by as
much as $17\%$.  Our constraint on
$M_{200,\mathrm{lens}}/M_{200,\mathrm{SZ}}$ should therefore be viewed
in the context of the cosmological model assumed in \citet{bleem15b},
from which our SZ-derived cluster masses are taken.  Additionally,
intrinsic scatter in the relationship between $M_{200,\mathrm{SZ}}$
and $M_{200,\mathrm{lens}}$ will lead to broadening of the likelihood
as a function of $M_{200,\mathrm{lens}}/M_{200,\mathrm{SZ}}$.
However, the expected level of intrinsic scatter between the true
cluster mass and $M_{200,\mathrm{SZ}}$ is only \,$\sim$15\% per
cluster \citep{benson13}.  Given our $3.1\sigma$ detection
significance across all clusters, the per-cluster constraint on the
lensing mass is roughly \,$\sqrt{513}/3.1 \sim 730\%$.  The effect of
intrinsic scatter in the SZ-derived masses is therefore only
$1-730/\sqrt{730^2 + 15^2} \sim 0.02\%$, and is therefore negligible
here.

\section{Conclusions}
\label{sec:conclusion}

We have presented a measurement of CMB cluster lensing using data from
the SPT.  Our data rule out the null hypothesis (that cluster lensing
is not occurring) at $3.1 \sigma$ and constrain the weighted average
cluster mass of our sample to be $M_{200} = 5.1_{-2.1}^{+2.5} \times
10^{14} \,\msun$ (68\% confidence limit).  Our cluster mass constraint
-- obtained by measurement of the CMB cluster lensing effect -- is
less precise than other cluster mass estimates, but it does offer a
confirmation of SZ-derived mass estimates with completely independent
sources of systematic errors:
${M_{200,\mathrm{lens}}}/{M_{200,\mathrm{SZ}}} = 0.83_{-0.37}^{+0.38}$
({\rm 68\%} \,C.L.).  Our lensing mass constraint is consistent with
${M_{200,\mathrm{lens}}}/{M_{200,\mathrm{SZ}}} = 1$ at $0.5\sigma$.

We have investigated several potential sources of systematic error and
have found that their individual effects are significantly less than
the statistical uncertainties of our mass constraints.  We find that
the most important systematic effects are the bulk velocity kSZ, the
kSZ due to a rotating cluster, lensing of foregrounds by the clusters,
cluster miscentering and deviation of the cluster density profile from
the NFW form in the outskirts of the cluster.  These findings are in
agreement with other investigations into CMB cluster lensing
systematic effects \citep[e.g.][]{Holder:2004,LewisKing:2006},
although the contaminating effects of foreground lensing appear to be
underappreciated in the literature.

All of the five most important systematic effects listed above bias
our lensing constraint to lower masses.  In our mock analysis, the
presence of these five systematic effects results in an average bias
of 39\% to lower cluster mass.  This level of bias amounts to roughly
$0.85\sigma$ in units of the statistical error bar.  We emphasize,
though, there are several uncertainties involved in the calculation of
this bias.  For one, we have almost certainly overestimated the
effects of foreground lensing on our analysis by placing all
foreground emission at $z = 4$.  Furthermore, our estimate of the bias
caused by cluster miscentering is likely an overestimate as well
because of our simplified treatment of this effect.  Finally, the
amplitude of the rotating-cluster kSZ signal is poorly constrained at
present, and its effects on our analysis are therefore somewhat
uncertain. Because of the large uncertainties associated with our
estimates of systematic effects, we have chosen to not include
corrections for these effects in our reported detection significance,
and instead compute the detection significance from the statistical
error bar alone.

Correcting for the measured 39\% bias to lower cluster mass would
cause the likelihood to prefer higher cluster mass and would therefore
yield a higher detection significance as well as a higher
${M_{200,\mathrm{lens}}}/{M_{200,\mathrm{SZ}}}$.  If the same bias is
assumed for each cluster, a 39\% shift to higher cluster mass would
cause the best fit ${M_{200,\mathrm{lens}}}/{M_{200,\mathrm{SZ}}}$ to
increase to roughly 1.15, still consistent with
${M_{200,\mathrm{lens}}}/{M_{200,\mathrm{SZ}}} = 1$ to within the
error bars.  There is therefore no evidence from this analysis of
tension with the SZ-derived cluster masses, even accounting for
potentially large systematic biases.

Additionally, as discussed in \S\ref{sec:spt}, systematic uncertainty
on $M_{200,\mathrm{SZ}}$ may affect our constraint on
${M_{200,\mathrm{lens}}}/{M_{200,\mathrm{SZ}}}$.  In particular, the
SZ-derived masses used in this work could potentially be overestimated
by as much as 8\% or underestimated by as much as 17\%, depending on
the assumed cosmological parameters.  Our constraint on
${M_{200,\mathrm{lens}}}/{M_{200,\mathrm{SZ}}}$ is derived assuming
the same cosmological parameters used in \citet{bleem15b} and should
be considered in that context.  Even if the maximal bias is assumed
for the SZ-derived cluster masses, our analysis does not yield tension
with ${M_{200,\mathrm{lens}}}/{M_{200,\mathrm{SZ}}} = 1$ at greater
than $1\sigma$.  This statement remains true even if the
lensing-derived masses are increased by 38\% to account for the
systematic biases discussed above.

Upcoming data sets offer the exciting possibility of significantly
improved measurements of CMB cluster lensing.  The measurement
presented here using data from the SPT-SZ survey is noise limited: the
lensing signal is at the few \muK\, level and is on few arcminute
scales, while the noise in the tSZ-free linear combination is roughly
55 $\mu$K-arcmin.  Ongoing experiments such as SPTpol
\citep{austermann12} and ACTPol \citep{naess14}, and future
experiments such as SPT-3G \citep{benson14}, Advanced ACTPol
\citep{Calabrese:2014}, the Simons Array \citep{Arnold:14}, and
so-called Stage IV CMB experiments
\citep[e.g.,][]{snowmass13neutrinos} will have significantly lower
noise levels than the SPT-SZ survey, allowing them to obtain
significantly stronger detections of the CMB cluster lensing signal.
Furthermore, these experiments will include additional information
about lensing in the form of polarization data.  In the primordial
CMB, the odd-parity (B-mode) component of the CMB polarization field
is expected to be uncorrelated with both the temperature field and the
even-parity (E-mode) component of the polarization field.
Consequently, lensing induced correlations between B modes and either
temperature modes or E modes can be used as a relatively clean probe
of CMB lensing \citep[e.g.,][]{hu02a}.  Furthermore, polarization
offers another handle on eliminating contamination from the SZ effect.
The polarized SZ effect (both thermal and kinematic) from clusters is
expected to be significantly smaller (i.e.,~less than 10-100 nK,
\citealt{carlstrom02}) than the unpolarized effect, so polarization
observations should offer a less-contaminated window into CMB cluster
lensing \citep[e.g.,][]{Holder:2004}.

With higher sensitivity data than that employed here, CMB cluster
lensing has the potential to provide powerful constraints on cluster
masses.  In principle, these mass constraints can be used to improve
cluster mass-observable relationships that are essential for using
clusters as cosmological probes.  However, our analysis of potential
contaminating effects in \S\ref{subsec:systematics_results} suggests
that there is still much work to be done in reducing systematic errors
associated with measurements of CMB cluster lensing.  Particularly
important are contamination from the kSZ effect, lensing of
foregrounds, and departure from the NFW profile at large radii.  In
principle, both the kSZ and lensing of the foregrounds can be modeled
and incorporated into the analysis to eliminate any bias that these
effects introduce.  However, uncertainty on the amplitude of the kSZ
and uncertainty on the foreground redshift distribution limits our
ability to accurately perform this modeling at present.  

\begin{acknowledgements}
The South Pole Telescope is supported by the National Science
Foundation through grant PLR-1248097.  Partial support is also
provided by the NSF Physics Frontier Center grant PHY-1125897 to the
Kavli Institute of Cosmological Physics at the University of Chicago,
the Kavli Foundation and the Gordon and Betty Moore Foundation grant
GBMF 947.  This research used resources of the National Energy
Research Scientific Computing Center, a DOE Office of Science User
Facility supported by the Office of Science of the U.S. Department of
Energy under Contract No. DE-AC02-05CH11231.  We acknowledge the use
of the Legacy Archive for Microwave Background Data Analysis
(LAMBDA). Support for LAMBDA is provided by the NASA Office of Space
Science.  This work was supported in part by the Kavli Institute for
Cosmological Physics at the University of Chicago through grant NSF
PHY-1125897 and an endowment from the Kavli Foundation and its founder
Fred Kavli.  The McGill group acknowledges funding from the National
Sciences and Engineering Research Council of Canada, Canada Research
Chairs program, and the Canadian Institute for Advanced Research.
M. Dobbs acknowledges support from an Alfred P. Sloan Research
Fellowship.  S.~Dodelson is supported by the U.S. Department of
Energy, including grant DE-FG02-95ER40896.  T.~de~Haan is supported by
a Miller Research Fellowship.  Cluster studies at SAO are supported by
NSF grant AST-1009649.
\end{acknowledgements}

\bibliography{cmb_cluster_lensing,/Users/ericbaxter/PAPERS/BIBTEX/spt}

\end{document}